\documentclass[twocolumn,showpacs,preprintnumbers,amsmath,amssymb]{revtex4}
\usepackage{amssymb}

\usepackage{graphicx}
\usepackage{dcolumn}
\usepackage{bm}
\usepackage{multirow}

\begin{document}


\title{Coarse-grained Simulations of Chemical Oscillation in a Lattice Brusselator System }
\author{Ting Rao$^1$}
\author{Zhen Zhang$^1$}
\author{Zhonghuai Hou$^{1,2}$}\email{hzhlj@ustc.edu.cn}
\author{Houwen Xin$^{1}$}%
\affiliation{%
\emph{\emph{$^1$Department of Chemical Physics, $^2$Hefei National
Laboratory for Physical Sciences at Microscale, University of
Science and Technology of China, Hefei, Anhui, 230026, People's
Republic of China}}}


\begin{abstract}
Accelerated coarse-graining (CG) algorithms for simulating
heterogeneous chemical reactions on surface systems have recently
gained much attention. In the present paper, we consider such an
issue by investigating the oscillation behavior of a two-dimension
(2D) lattice-gas Brusselator model. We have adopted a coarse-grained
Kinetic Monte Carlo (CG-KMC) procedure, where $m \times m$
microscopic lattice sites are grouped together to form a CG cell,
upon which CG processes take place with well-defined CG rates. We
find that, however, such a CG approach almost fails if the CG rates
are obtained by a simple local mean field ($s$-LMF) approximation,
due to the ignorance of correlation among adjcent cells resulted
from the trimolecular reaction in this nonlinear system. By properly
incorporating such boundary effects, we thus introduce the so-called
$b$-LMF CG approach. Extensive numerical simulations demonstrate
that the $b$-LMF method can reproduce the oscillation behavior of
the system quite well, given that the diffusion constant is not too
small. In addition, we find that the deviation from the KMC results
reaches a nearly zero minimum level at an intermediate cell size,
which lies in between the effective diffusion length and the minimal
size required to sustain a well-defined temporal oscillation.

\end{abstract}

\pacs{05.10.Ln, 82.40.Bj, 02.70.Uu}

\maketitle
\section{Introduction}
When driven far from thermal equilibrium, heterogeneous surface
chemical reaction systems often show a variety of complex
dissipative structures such as oscillations, Turing patterns, spiral
waves and turbulence\cite{A.S.M and G.E, Mik4, Mik5,
Progress-165,Science-293,Nature-V390, PRL-V65,Science-278, JCS,
Book}. Traditionally, these phenomena are mainly observed at
macroscopic scales ranging from a few to several hundred
micrometers, but recent studies showed that they are also present in
nanoscale systems. At the present time, two different theoretical
approaches are used to describe these nonlinear behaviors of surface
reactions. Mean field deterministic equations (MDFE), like the
reaction-diffusion equation, can provide good qualitative
description of spatiotemporal dynamics\cite{MF1,MF2}. However, they
are essentially phenomenological and neglect microscopic mechanisms
such as lateral interactions between adsorbate molecules. In
addition, MFDE also ignores the molecular fluctuations which may
play important roles in mesoscopic systems. Another approach,
microscopic lattice models, takes explicitly into account the
adsorption, desorption, diffusion and reaction processes as random
events, and one can use kinetic Mente Carlo (KMC) methods to yield
detailed valuable information about the microscopic reaction
properties\cite{KMC1, KMC2, KMC3, KMC4, KMC5, KMC6}. This
microscopic method can account for the molecular interactions and
fluctuations directly, however, memory and speed of available
computers limit the maximal spatial size of the system, which
renders the direct KMC simulation of nanoscale spatiotemporal
structures and populations a difficult task. Therefore, a promising
way is to develop coarse-grained (CG) approaches, bridging the gap
between those two, aiming at significantly reducing the degree of
freedom to accelerate the simulation on large length scale while
properly preserving the microscopic fluctuation information and
correct dynamics.

Very Recently, two kinds of CG methods based on lattice-gas model
have been proposed. One is the continuum mesoscopic modeling
devoloped by Mikhailov et.al,\cite{Mik0, Mik1, Mik2, Mik3, Mik6,
Mik7} which is derived from coarse-graining of the underlying
microscopic master equation to get the functional Fokker-Planck
equation and its corresponding stochastic partial differential
equations (SPDE). The other is a discrete CG approach proposed by
Vlachos and coworkers \cite{{Vlachos0},{Vlachos1},{Vlachos2},
{Vlachos3}, {Vlachos4},{Vlachos6}, {Vlachos7}, {Vlachos8}}, which is
a kind of CG-KMC algorithm by grouping the microscopic lattice sites
into coarse cells and the CG system evolves by a sequence of CG
events associated with the microscopic processes. By numerically
solving the SPDE, Mikhailov has successfully investigated the
nucleation of single reactive adsorbate in one-dimension (1D) and 2D
systems \cite{Mik0}, the formation of stationary microstructures for
single species with attractive lateral interactions in a 2D
system\cite{Mik1} and the formation of traveling nanoscale
structures in a model of two different species on 2D reaction
surface \cite{Mik2}. Vlachos et.al had in-depthly investigated the
validity of the CG-KMC approach in 1D conceptional systems with
different potential form \cite{{Vlachos1},{Vlachos2}}, the 1D Ising
system with spin exchange \cite{Vlachos3}, prototype model of 1D
diffusion through a membrane\cite{Vlachos4}, steady pattern
formation of simple reaction model on 2D surface\cite{Vlachos6}, and
so on. Furthermore, the authors had also discussed the possibility
for the extension of CG-KMC to complex lattices, multicomponent
systems \cite{Vlachos7} and heterogeneous plasma membranes
\cite{Vlachos8}. It has been shown that both CG methods enable
dynamic simulations over large length and time scales and can
accurately capture transient and equilibrium solutions as well as
noise properties especially for long-ranged potentials.

In the present paper, our mainly purpose is to find an effective
CG-KMC method to investigate the nonlinear oscillation behaviors of
surface chemical system which has already been developed into a
field of very active research \cite{G Ertl, Book}. Here, we adopt a
2D lattice-gas Brusselator model, which is a typical oscillatory
system with a nonlinear autocatalytic trimolecular reaction. To
preserve the microscopic information correctly and accelerate the
simulation at the same time, a CG-KMC algorithm by grouping $m
\times m$ microscopic sites into a CG cell is adopted. Numerical
results show that such a CG method is actually not a good
approximation if the CG rates are obtained by a simple local mean
field ($s$-LMF) approximation, due to the correlation among adjacent
cells resulted from the trimolecular reaction cannot be neglected.
We thus proposed a $b$-LMF approach, which has properly accounted
for the boundary corrections to the CG reaction rates. Extensive
numerical simulations show that the $b$-LMF method can reproduce
quite well the oscillation behaviors, obtained from the KMC
simulations on the microscopic lattice. To quantitatively
investigate the accuracy of the CG methods, we introduce deviation
coefficients $\gamma_A$ for the oscillation amplitude and $\gamma_T$
for the oscillation period between the CG-KMC and KMC, respectively.
We find again that the $b$-LMF is remarkably better than $s$-LMF,
and there is an intermediate CG cell size where the deviation
reaches a nearly zero minimal level. We suggest that for the CG
method to work, the cell size should lie in between the effective
diffusion length and the minimal size required to sustain a
well-defined temporal oscillation.

The paper is organized as follows. In Section \ref{sec2}, we present the lattice Brusselator model and describe the methods in detail, including the KMC and CG-KMC procedures. The numerical results are given in \ref{sec3}, where we mainly focus on the comparison between KMC and CG-KMC, by investigating the deviations in the oscillation amplitude and period as functions of the control parameter. We end by conclusions in \ref{sec4}.

\section{Model and Method}
\label{sec2}
\subsection{The Brusselator Model}
We consider a modified Brusselator model on the 2D surface lattice as follows:
\begin{align}
U_g + *  &\xrightarrow{\text \em K_1} U_a  \label{eq1} \\
U_a &\xrightarrow{\text \em K_2} * + U_g \label{eq2} \\
U_a &\xrightarrow{\text \em K_3} V_a \label{eq3} \\
V_a &\xrightarrow{\text \em K_4} * + V_g \label{eq4} \\
2U_a + V_a &\xrightarrow{\text \em K_5} 3U_a \label{eq5} \\
U_a + * &\xrightarrow{\text \em K_6} * + U_a \label{eq6} \\
V_a + * &\xrightarrow{\text \em K_7} * + V_a \label{eq7}
\end{align}
Herein, the reactions under consideration are assumed to run on a $N
\times N$ square lattice. Sites are either vacant (denoted by * ) or
occupied by single $U$ or $V$ particles(See Fig. 1$a$). The
subscripts 'g' and 'a' represent the species in gas phase and
adsorbed on the surface, respectively. Process (1) denotes the
adsorption of species $U$, (2) the desorption of $U$, (3) the
conversion from adsorbed $U$ to $V$, and (4) the desorption of $V$,
respectively. Step (5) is the autocatalytic reaction wherein an
adsorbed $V$ molecule with two nearest neighbor $U$ molecules
converts to $U$. The parameters $K_{\alpha} (\alpha=1,...,5)$
represent dimensionless rate constants. Steps (6) and (7) denote the
diffusion processes of $U$ and $V$, respectively, where $K_6$ and
$K_7$ are the corresponding diffusion constants. In Table I, these
processes and their corresponding propensity functions are listed.
Herein, $w_{i\alpha}$ ($\alpha=1,...,7$) represent the propensity
function of the $\alpha$-th process taking place at site $i$. The
occupation function $\sigma_i^\phi$ (where $\phi=U$ or $V$) denotes
the state of a given surface site $i$: $\sigma_i^\phi=1$ if site $i$
is occupied by species $\phi$ and 0 otherwise. Note that a vacant
site is necessary for the adsorption of $U$ or the diffusion
processes. In the reaction process (5), both sites $j$ and $k$ must
be nearest neighbors of site $i$.

\begin{table*} \caption{Stochastic processes and corresponding reaction rates in KMC simulation for the lattice-gas Brusselator model.}
\renewcommand{\baselinestretch}{1.75}\footnotesize 
\setlength{\tabcolsep}{2pt}
\begin{center}
\begin{tabular}{ccc}
\hline \hline Process Description & State Change &KMC Rate  \\
 \hline U Adsoprtion & $\sigma^U_i: 0\rightarrow 1$ &${w}_{i1} = K_1(1-\sigma^{U}_i-\sigma^{V}_i)$ \\
U Desoprtion & $\sigma^U_i: 1\rightarrow 0$ &${w}_{i2} = K_2\sigma^{U}_i$ \\
U Conversion to V & $\sigma^U_i: 1\rightarrow 0$, $\sigma^V_i: 0\rightarrow 1$ &${w}_{i3} = K_3\sigma^{U}_i$ \\
V Desorption & $\sigma^V_i: 1\rightarrow 0$ &${w}_{i4} = K_4\sigma^{V}_i$ \\
2U+V Reaction & $\sigma^V_i: 1\rightarrow 0$, $\sigma^U_i: 0\rightarrow 1$ & ${w}_{i5} = K_5\sigma^{V}_i\sigma^{U}_j\sigma^{U}_k$ \\
U Diffusion & $\sigma^U_i: 1\rightarrow 0$, $\sigma^U_j: 0\rightarrow 1$ &${w}_{i6} = K_6\sigma^{U}_i(1-\sigma^{U}_j-\sigma^{V}_j)$ \\
V Diffusion & $\sigma^V_i: 1\rightarrow 0$, $\sigma^V_j:
0\rightarrow 1$ &${w}_{i7}
=K_7\sigma^{V}_i(1-\sigma^{U}_j-\sigma^{V}_j)$ \\
 \hline \hline
\end{tabular}
\end{center}
\end{table*}

\subsection{MF Description}
Assuming that the surface is well-mixed by the diffusion process,
one may describe the system dynamics by the following MF
deterministic equations,
\begin{align}
 \frac{du}{dt} = (w_1-w_2-w_3+w_5), \frac{dv}{dt} = (w_3-w_4-w_5),
\end{align}
where $u$ and $v$ are respectively the surface coverage of species U and V.
$w_{\alpha=1,...,5}$ denote the rate of reaction-$\alpha$ with
\begin{align}
w_1&={K}_1(1-u-v), \ w_2=K_2 u, w_3=K_3 u, \nonumber\\  w_4&=K_4 v,
w _5=K_4 vu^2.
\end{align}
In the present work, we use this MF equation to determine the
bifurcation diagram of the system. In certain parameter ranges, the
system can undergo a Hopf bifurcation, where stable limit cycle
emerges.

The MF equation does not take into account internal fluctuations
inherent in chemical reaction systems. For small systems, such
fluctuations may play important roles. To account for the internal
noise while keep the MF approximation, one may use the chemical
Langevin equations as follows,
\begin{align}
 \frac{du}{dt} = &(w_1-w_2-w_3+w_5) \nonumber \\
               &+\frac{1}{N}[\sqrt{w_1}\xi_1(t)-\sqrt{w_2}\xi_2(t)-\sqrt{w_3}\xi_3(t)+\sqrt{w_5}\xi_5(t)]\\
 \frac{dv}{dt} = &(w_3-w_4-w_5) \nonumber\\
                 &+\frac{1}{N}[\sqrt{w_3}\xi_3(t)-\sqrt{w_4}\xi_4(t)-\sqrt{w_5}\xi_5(t)]
\end{align}
where $\xi_{\alpha=1,...,5}(t)$ are Gaussian white noises associated
with the reaction channels, obeying $<\xi_{\alpha}(t)>=0$ and
$<\xi_{\alpha}(t)\xi_{\beta}(t')>=\delta_{\alpha\beta}\delta(t-t')$.
The items in the bracket with $\xi_{\alpha}(t)$ give the internal
noises, scaling as $1/N$, which are closely coupled with the
reaction kinetics. In the macroscopic limit $N \rightarrow \infty$,
the internal noise items can be ignored and the CLE recovers the
deterministic equation \textbf{(8)}.

\subsection{KMC Simulations }
Given the processes and their propensity functions as listed in
Table I, one can then perform KMC to study the dynamics. In the
present work, we adopt a null-event KMC procedure. The main steps
can be outlined as follows,
\begin{enumerate}
  \item Determine which process $\alpha$ to happen. To do this, we first draw a random number $r_1$ from the uniform
distribution in the unit interval, and then take $\alpha$ as the
smallest integer satisfying $\sum^\alpha_{\beta=1} K_\beta > r_1
W_0$, where $W_0=\sum^7_{\beta=1} K_\beta $ denotes the total
maximum transition rates.
  \item Randomly select a surface site $i$ with equal probability $p_i = 1/N^2$.
  \item Determine whether the selected process $\alpha$ can take place on site $i$ or not. This is given by a so-called
participation index \cite{Review of KMC}
$\epsilon_{i\alpha}=w_{i\alpha}/K_\alpha$, which takes value 0 or 1,
corresponding to the propensity functions $w_{i\alpha}$ shown in
Table I. Note that this index depends on the local configuration
around site $i$ and varies with time. For the desorption process
(2), for instance, $\epsilon_{i2}=1$ if site $i$ is occupied by $U$
and 0 otherwise. For the diffusion of $U$(or $V$),
$\epsilon_{i\alpha}=1$ if the site $i$ is occupied by $U$ (or $V$)
and a randomly select nearest neighbor $j$ is vacant. For the
reaction process (5), the index reads 1 if site $i$ is $V$, and two
succeedingly selected nearest neighbors $j$ and $k$ are both $U$.
Note here we have used the same rule for this process as that
proposed by Zhadnov \cite {Zhadnov2001}: As shown in Fig.1, reaction
can only happen among orthogonal configurations, such as those shown
by (a), (b) and (c), but not within line configurations such as (d).

  \item Execute the process $\alpha$ if $\epsilon_{i\alpha}=1$, and terminate the trial otherwise.
  \item Repeat steps 1 to 4.
\end{enumerate}

In the present work, we  start the KMC run from a clean surface. The
time advancement can be measured in terms of $\tau_{KMC}=1/W_0$. The
KMC results are assumed to be correct, and we use them to check the
validity of other methods, especially that of the CG-KMC.

\begin{figure}
\begin{center}
\includegraphics [width=8cm]{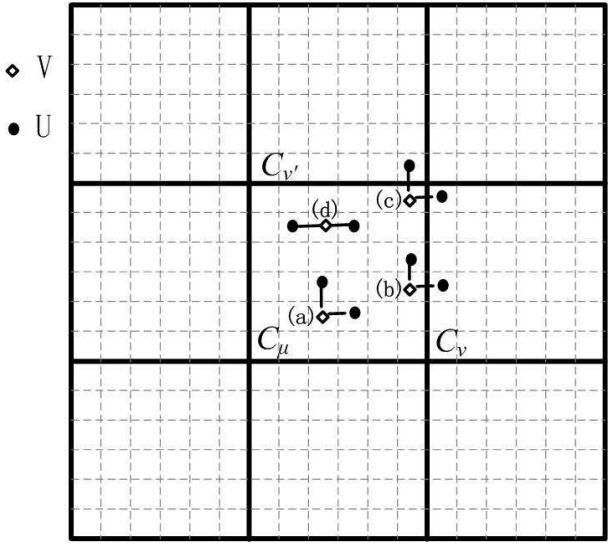} \caption{The scheme of spatial coarse graining is plotted. Here,
a $18 \times 18$ microscopic lattice (see dash-line) is divided into
a $3\times3$ CG lattice(see solid-line) by uniformly grouping
$6\times6$ microscopic sites to a CG cell. At the microscopic level,
the reaction $2U+V\rightarrow 3U$ can only take place on orthogornal
UVU configurations (a), (b) and (c), but not on line configuration
(d). \label{fig1}}
\end{center}
\end{figure}

\subsection{The CG-KMC Methods}
In the present work, we adopt a CG procedure originally introduced
by Vlachos et.al \cite {Vlachos3}. Since the diffusion processes are
usually faster than other slow processes, it is reasonable to assume
that particles are well-mixed in a comparatively large domain, whose
scale is determined by the diffusion length, over short time scales.
Therefore, one can divide the micro-lattice into several coarse
cells, wherein each particle is assumed to have an equal probability
of occupying any microscopic lattice site, and no spatial
correlation exists. Obviously, the most natural way for such a
spatial coarse graining on 2D surface is to group $m \times m$ sites
into a CG cell. For instance, Fig.1 shows the coarse-graining of a
$18 \times 18$ micro-lattice into a $3\times 3$ CG-lattice, wherein
each CG-cell contains $m^2=36$ sites (hereafter, we use 'site' for
the micro-lattice and 'cell' for the CG-lattice).

To perform CG-KMC, one needs to define the CG processes taking place
on the CG lattice and obtain the corresponding CG rates. We now
introduce the CG variables $$\eta^{\phi}_\mu = \sum_{i\in C_\mu }
\sigma^\phi_i \ \ \ \text{and} \ \ \ \bar{\eta}^{\phi}_\mu =
\eta^{\phi}_\mu/m^2 $$ to denote the number and coverage of
$\phi$-species in the $\mu$-th CG-cell $C_\mu$, respectively.
Consider the micro-process (1), for example, the CG process is
defined as the adsorption of one $U$ particle in any site $i$ inside
a CG-cell $C_\mu$. Since $U$-adsorption only involves one surface
site, there is no spatial correlation between different adsorption
events inside a cell, therefore the rate of the CG adsorption
process easily reads
    $$w_{\mu 1}=\sum_{i\in C_\mu} K_1(1-\sigma_i^U-\sigma_i^V)=m^2 K_1(1-\bar{\eta}^U_\mu-\bar{\eta}^V_\mu).$$
Following this simple rule, one can readily obtain the CG rates for
the single-site processes (1) to (4).

For the diffusion processes (6) and (7), one must bear in mind that
the diffusion constant between two adjacent CG-cell, $\tilde K_6$ (
$\tilde K_7$ ), is not identical to that between two adjacent
micro-sites, $K_6(K_7)$. To establish the relationship between these
two rates, we can adopt the so-called 'flux-consistency' rule, which
requires that the average flux across the boundary of two adjacent
CG cells, calculated from the CG diffusion process, should be the
same as that calculated from averaging over the micro-diffusion
process. By using this criterion and considering the maintenance of
detail-balance, see \cite{Vlachos2}, one must have $\tilde
K_6=K_6/m^2$ and $\tilde K_7=K_7/m^2$. Finally, the rate for the CG
diffusion process of $U$ is
$$w_{\mu 6}=\sum_i \tilde K_6 \langle\sigma_i^U (1-\sigma_j^U-\sigma_j^V) \rangle_{i\in C_\mu,j\in C_\nu} = \frac{K_6}{m^2}
\eta_\mu^U (1-\bar \eta_\nu^U-\bar \eta_\nu^V). $$ Herein, $\langle
\cdot \rangle$ means ensemble average. In the final equation, we
have ignored the correlation between different cells and simply
replaced the ensemble average of $\sigma_i^\phi$ inside $C_\mu$ by
$\bar \eta_\mu^\phi$ \cite{Vlachos1, Vlachos2}. The rate for CG
diffusion of $V$ can be obtained in a similar way.

For the trimolecular reaction process (5), however, strong
correlation exists between neighboring sites. To perform the CG
simulation, we need to express the ensemble-averaged rate of this
process inside a cell by a function of the CG-variables,
   $$w_{\mu 5}=F(K_5, m,  \bar \eta_\mu^U, \bar \eta_\mu^V).$$
However, it is hard to decide the correct functional form at this
stage. It is worthy to note here that a seamless approach has been
proposed very recently to address the validity of reaction-diffusion
master equations\cite{PNAS}. The authors argued that to make the
master equation to be consistent with the micro-model, the reaction
constant must be dependent on the coarse-size, here is $m$. They
demonstrated the success of this idea for a reversible
aggregation-dissociation reaction. Unfortunately, it is rather
difficult for us to work out a similar result for the trimolecular
lattice gas system considered here. Therefore, to step forward, we
have to make some approximations. To the lowest order, one may use
the simple local mean field (s-LMF) approximation as follows,
$$w_{\mu 5}=m^2 \langle K_5 \sigma_i^V \sigma_j^U \sigma_k^U \rangle'_{i,j,k\in C_\mu} = m^2 K_5 \bar \eta_\mu^V
(\bar\eta_\mu^U)^2.$$ Herein, $\langle \cdot \rangle'$ denotes the
ensemble average over the orthogonal configurations inside $C_\mu$.
Given that the diffusion process is fast and the cell size is
smaller than the diffusion length, this approximation, although
crude, might be acceptable.

Unfortunately, as we will show(see below), however,  this s-LMF
scheme for the reaction process almost fails to reproduce the KMC
results, no matter how large the coarse-size $m$ and the diffusion
constants $K_6(K_7)$ are. It seems that the s-LMF loses some key
components that should be considered during the CG procedure. We
note here that the s-LMF scheme totally ignores the correlations
between adjacent cells resulted from the trimolecular process. For
instance, the reaction configuration (b) (see in Fig.1) on the
border of cell $\mu$ involves two sites ($U$ and $V$) in cell $\mu$
and one site ($U$) in cell $\nu$. If we use LMF for both cell $\mu$
and $\nu$, the ensemble averaged rate for this particular
configuration should read $ K_5 \bar \eta^U_\mu \bar \eta^V_\mu \bar
\eta^U_\nu$, which is different from the rate for the reaction
inside $C_\mu$, $ K_5 (\bar \eta^U_\mu)^2 \bar \eta^V_\mu$, if we
consider that concentration gradients exist between adjacent cells.
We argue that this effect should be taken into account to compensate
the discrepancy resulted from the CG approximation. Similarly, the
reaction configuration (c)(in Fig. 1) at the corner of cell $\mu$
involves sites in three adjacent cells. Therefore, instead of the
s-LMF, one may use a boundary-corrected LMF (b-LMF) scheme by
writing down the CG-rate of process (5) as follows,
\begin{equation}
 w^{b}_{\mu5} = f_1\tilde w_{intra} + f_2 \tilde w_{border} + f_3\tilde w_{corner}.
\label{eq12}
\end{equation}
Herein,
   \begin{equation}
     \tilde w_{intra} = m^2 K_5 \bar \eta^V_\mu (\bar \eta^U_\mu)^2 ,
   \end {equation}
   \begin{equation}
     \tilde w_{border} = m^2 K_5 \bar \eta^V_\mu \bar \eta^U_\mu (\sum_\nu \bar \eta^U_\nu/4),
   \end {equation}
and
   \begin{equation}
     \tilde w_{corner} = m^2 K_5 \bar \eta^V_\mu (\sum_{\nu \nu'} \bar \eta^U_\nu \bar \eta^U_{\nu'}/4)
   \end {equation}
denote the average rate of process (5) inside, on the border of, and
at the corner of cell $\mu$, respectively.  Note that the summation
in equation (14) runs over the four adjacent cells of $C_\mu$, and
that in (15) runs over adjacent cells of the four corners. The
weighting factors $f_1$, $f_2$ and $f_3$ denote the possibility of
finding an reaction $UVU$ configuration belonging to the three
categories, respectively, given that the $V$ site is inside the
current cell $C_\mu$. By simple manipulations, we have
$$f_1=(1-1/m)^2, \ f_3=1/m^2, \ \ \texttt{and}\ \  f_2=1-f_1-f_3.$$

\begin{table*} \caption{CG processes and rates associated with the CG cell $C_\mu$.}
\begin{center}
\footnotesize
\begin{tabular}{cccc}
\hline\hline Process Description & State Change &$s$-LMF Rate &$b$-LMF Rate\\
\hline U Adsorption &$\eta^U_\mu \rightarrow \eta^U_\mu+1$ &${
w}_{\mu1} = m^2 {K}_1
 (1- \bar {\eta}^{U}_\mu- \bar {\eta}^{V}_\mu)$ &${w}_{\mu1}$ \\
U Desorption & $\eta^U_\mu \rightarrow \eta^U_\mu-1$ &$w_{\mu2} =  m^2 {K}_2 \bar{\eta}^{U}_\mu$ &$w_{\mu2}$ \\
U Conversion to V &$\eta^U_\mu \rightarrow \eta^U_\mu-1$, $\eta^V_\mu\rightarrow \eta^V_\mu+1$ &$w_{\mu3} =  m^2 {K}_3 \bar{\eta}^{U}_\mu$ &$w_{\mu3}$ \\
V Desorption & $\eta^V_\mu\rightarrow \eta^V_\mu-1$ &$w_{\mu4} = m^2{K}_4 \bar{\eta}^{V}_\mu$ &$w_{\mu4}$ \\
2U+V Reaction &$\eta^V_\mu\rightarrow \eta^V_\mu-1$, $\eta^U_\mu\rightarrow \eta^U_\mu+1$ &$w_{\mu5}=m^2{K}_5 \bar{\eta}^{V}_\mu (\bar\eta_\mu^U)^2$ &$w_{\mu5}^{b}$\\
U Diffusion &$\eta^U_\mu \rightarrow \eta^U_\mu-1$, $\eta^U_\nu
\rightarrow \eta^U_\nu+1$ &$w_{\mu6} = {K}_6
 \bar{\eta}^{U}_\mu
 (1-\bar{\eta}^{U}_\nu-\bar{\eta}^{V}_\nu)$ &$w_{\mu6}$ \\
V Diffusion & $\eta^V_\mu \rightarrow \eta^V_\mu-1$,$\eta^V_\nu
\rightarrow \eta^V_\nu+1$ &$w_{\mu7} = {K}_7 \bar{\eta}^{V}_\mu
 (1-\bar{\eta}^{U}_\nu-\bar{\eta}^{V}_\nu)$ & $w_{\mu7}$ \\
\hline \hline
\end{tabular}
\end{center}
\end{table*}

In Table II, the CG processes as well as their corresponding CG
rates are listed. Note that b-LMF and s-LMF show difference only for
the trimolecular reaction. According to these processes and rates,
one can readily perform CG-KMC simulations. In the present paper, we
also use null-event procedure as that for the KMC. The steps are
outlined as follows,
\begin{enumerate}
\item Choose a process similar to the first step used in the KMC, except that now $K_6$ and $K_7$ should be replaced by
$\tilde
K_6=K_6/m^2$ and $\tilde K_7=K_7/m^2$. Correspondingly, $W_0$ should be changed to $W^{CG}_0$.
\item Randomly select a cell $\mu$ with equal probability.
\item Calculate the reaction probability $\epsilon_{\mu\alpha}$ for the process $\alpha$ to happen associated with the current
cell $\mu$. This probability simply equals to $w_{\mu \alpha}/(K_\alpha m^2)$ for $1 \le \alpha \le 5$ and $w_{\mu
\alpha}/(\tilde K_\alpha m^2)$ for $\alpha=6$ or 7.
\item Generate a second uniformly distributed random number $r_2$ in the unit interval. If $r_2 \le \epsilon_{\mu\alpha}$,
execute the process $\alpha$, and the trial ends otherwise.
\item Repeat the above steps.
\end{enumerate}

In the present study, we start CG-KMC simulations from the same
initial conditions as in the KMC. To be consistent with the KMC, the
time increment should read $\tau_{CG}=1/W^{CG}_0$ for each trial. We
compare the CG-KMC results with the KMC ones to check the validity
of CG approaches.

\section{Numerical Simulations and Discussion}
\label{sec3} In our work, the main parameters used in simulations
are $K_1=5.0\times10^{-5}$, $K_2=1.0\times10^{-3}$,
$K_3=5.0\times10^{-3}$ and $K_4=6.0\times10^{-5}$, while $K_5$ and
$K_6=K_7=D$ are control parameters. To compare the results of
different methods, we have generated time series $u(t)$ or $v(t)$
with enough length and calculated the oscillation amplitude $A$ and
period $T$ as a function of the control parameters. In Fig.(2a), the
dependence of the oscillation range of $v$ on $K_5$ is shown,
obtained by the MFDE,  CLE and KMC with different diffusion constant
$D$. Correspondingly the curves for the period $T$ are drawn in
Fig.(2b). The CLE results are obtained by numerical simulation of
Eq.(10) and (11) with a time step $\Delta t=0.01$ and $N=256$. All
the KMC results are also performed on a $256\times 256$ square
lattice.  The solid line obtained from MFDE corresponds to the
bifurcation diagram. A Hopf bifurcation locates at $K_{5c} \simeq
2.15$, below which deterministic oscillation can be observed.
Several points can be addressed from this figure. First of all, CLE
and KMC show strong qualitative differences with the MFDE:
Stochastic oscillations can be observed even outside the
deterministic oscillatory region, here $K_5>K_{5c}$. This so-called
noise induced oscillation phenomenon has gained great attention in
recent years and may have important applications especially in
circadian oscillation systems. Secondly, the KMC results depend
strongly on the diffusion constant $D$. However, the results for
$D=10$ and $D=30$ nearly collapse, indicating that the KMC results
may converge in the limit of large $D$. In this latter case, the
MFDE can reproduce the KMC results when the parameter lies deep
inside the oscillatory region, see the range $K_5<2.1$. If $D$ is
small, both MFDE and CLE show large discrepancies with the KMC
results, no matter the range of the control parameter. Finally, we
would like to point out here that the CLE cannot reproduce the KMC
results accurately even for large $D$, although they share some
qualitative features, e.g., noise induced oscillation to the right
side of the Hopf point.

\begin{figure}
\begin{center}
\includegraphics [width=0.45\columnwidth,height=5cm]{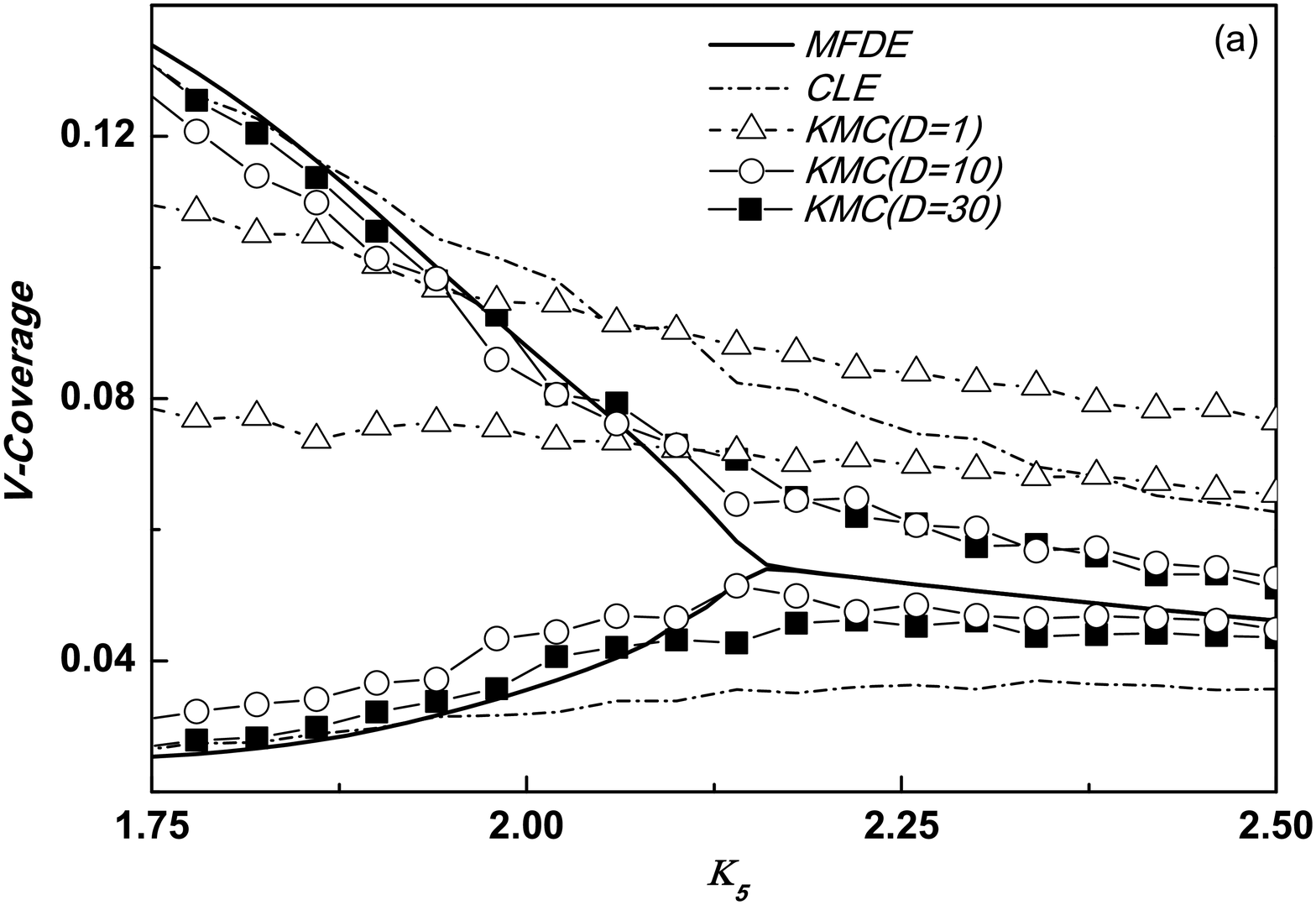}
\includegraphics [width=0.45\columnwidth,height=5cm]{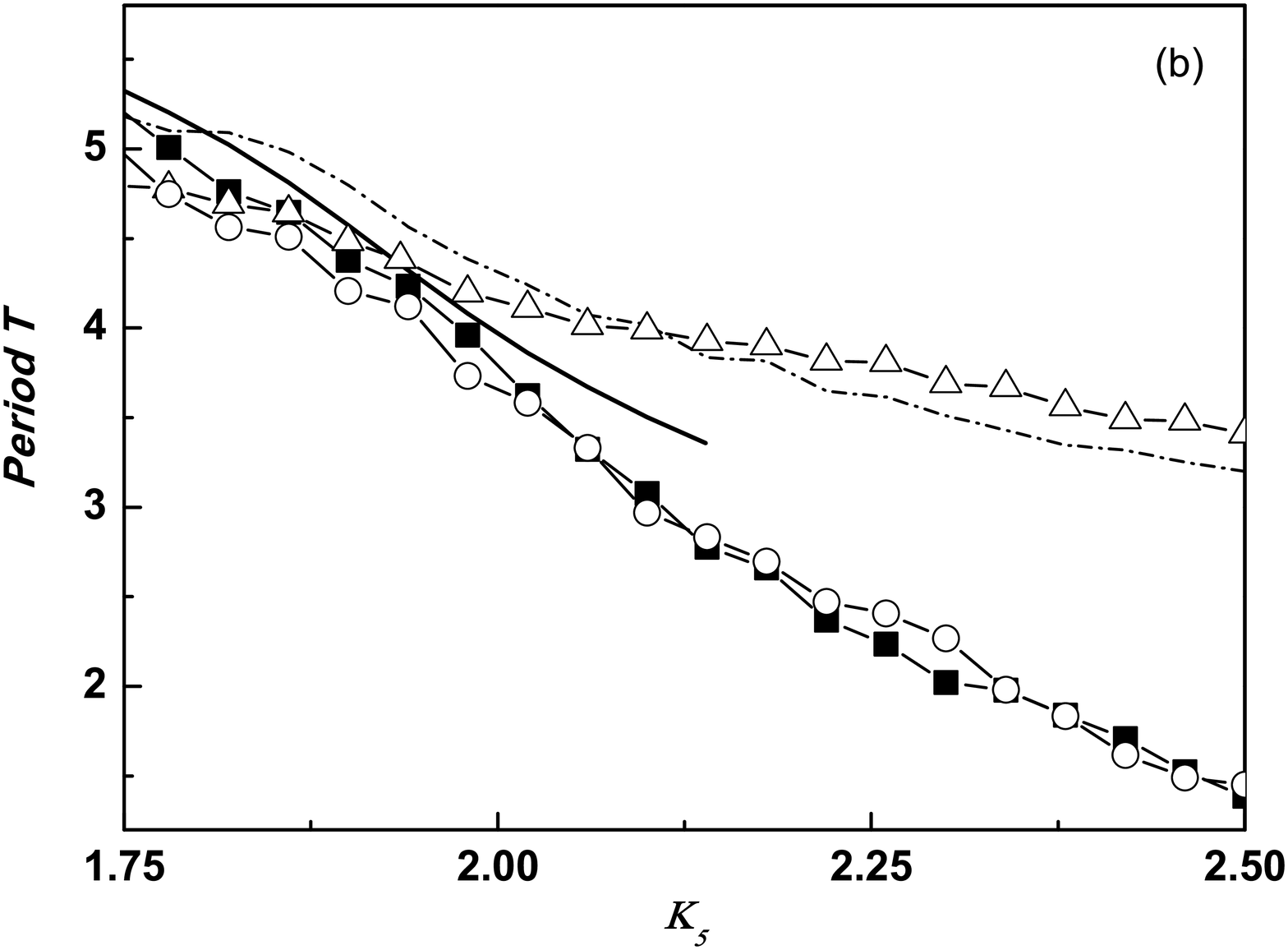}
\caption{The oscillation range of (a) $V$-coverage  and (b) period
are presented as functions of control parameter $K_5$, obtained by
MFDE, CLE and KMC with $D=1, 10, 30$, respectively. Parameters are
$K_1=5.0\times10^{-5}$, $K_2=1.0\times10^{-3}$,
$K_3=5.0\times10^{-3}$, $K_4=6.0\times10^{-5}$ and $N=256$.
\label{fig2}}
\end{center}
\end{figure}

\begin{figure}
\begin{center}
\includegraphics [width=0.45\columnwidth,height=5cm]{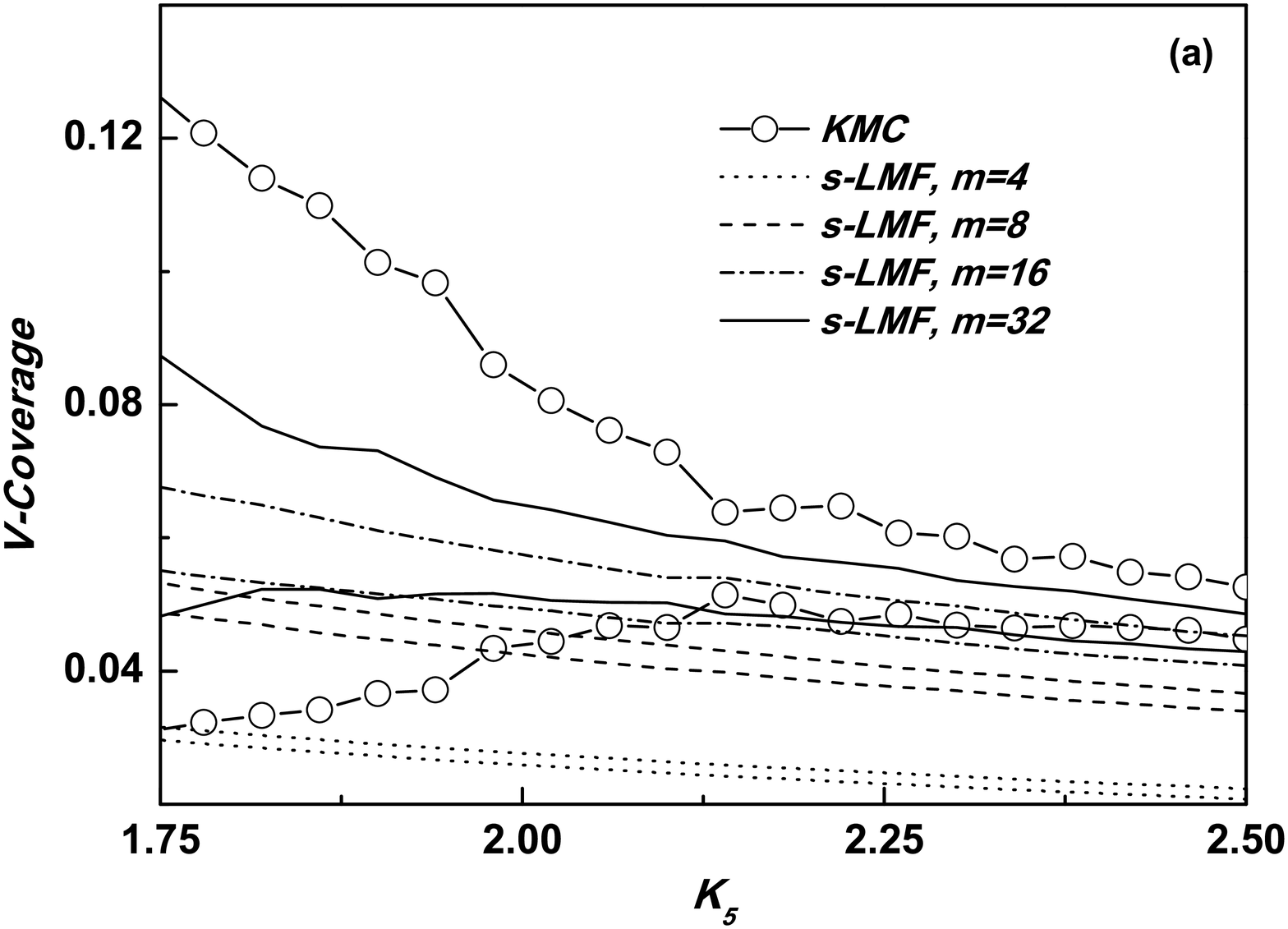}
\includegraphics [width=0.45\columnwidth,height=5cm]{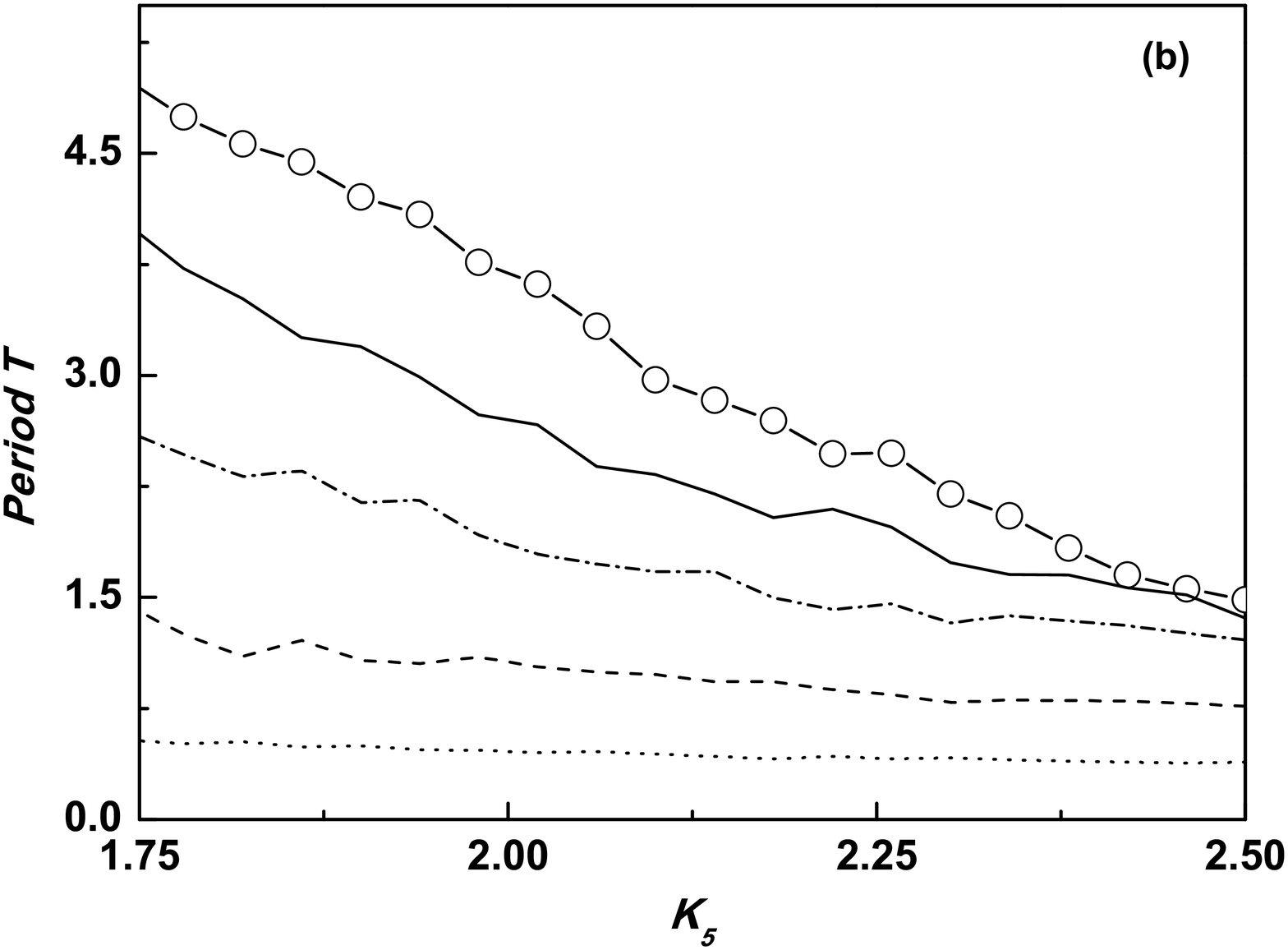}
\caption{ The oscillation range of (a) $V$-coverage  and (b) period
obtained from the $s$-LMF CG approach with different coarse size
$m$. The diffusion constant is $D=10$ and the other parameters are
the same as in Fig. 2. \label{fig3}}
\end{center}
\end{figure}

In the following part, we mainly consider a system with size $N=256$
and diffusion constant $D=10$. We have also performed some KMC
simulations on larger systems, e.g., $N=512$, but the main
conclusions are the same. Since we are mainly interested in the
validity of CG methods and extensive simulations are required to
compare the results of different methods, we have fixed $N=256$
throughout the paper. CG-KMC simulations are performed according to
the CG processes and rates listed in Table II. In Fig.3, the
oscillation amplitude and period obtained by using s-LMF rates are
shown, for different sizes of the CG cell. Apparently, the s-LMF
approach almost fails to reproduce the results of KMC, even
qualitatively. For small $m$, the s-LMF totally loses the whole
bifurcation features of the KMC dynamics. We show in Fig.4, however,
that the b-LMF behaves much better than the s-LMF. Firstly, the
b-LMF can reproduce the global bifurcation feature quite well, even
for small $m$. In addition, for an intermediate value of $m$, say,
$m=8$ here, the b-LMF results show excellent consistent with the KMC
results, in both the oscillation amplitude (Fig.4a) and the period
(Fig.4b). It seems that the b-LMF approach does catch some key
factors during the CG procedure.

\begin{figure}
\begin{center}
\includegraphics [width=0.45\columnwidth,height=5cm]{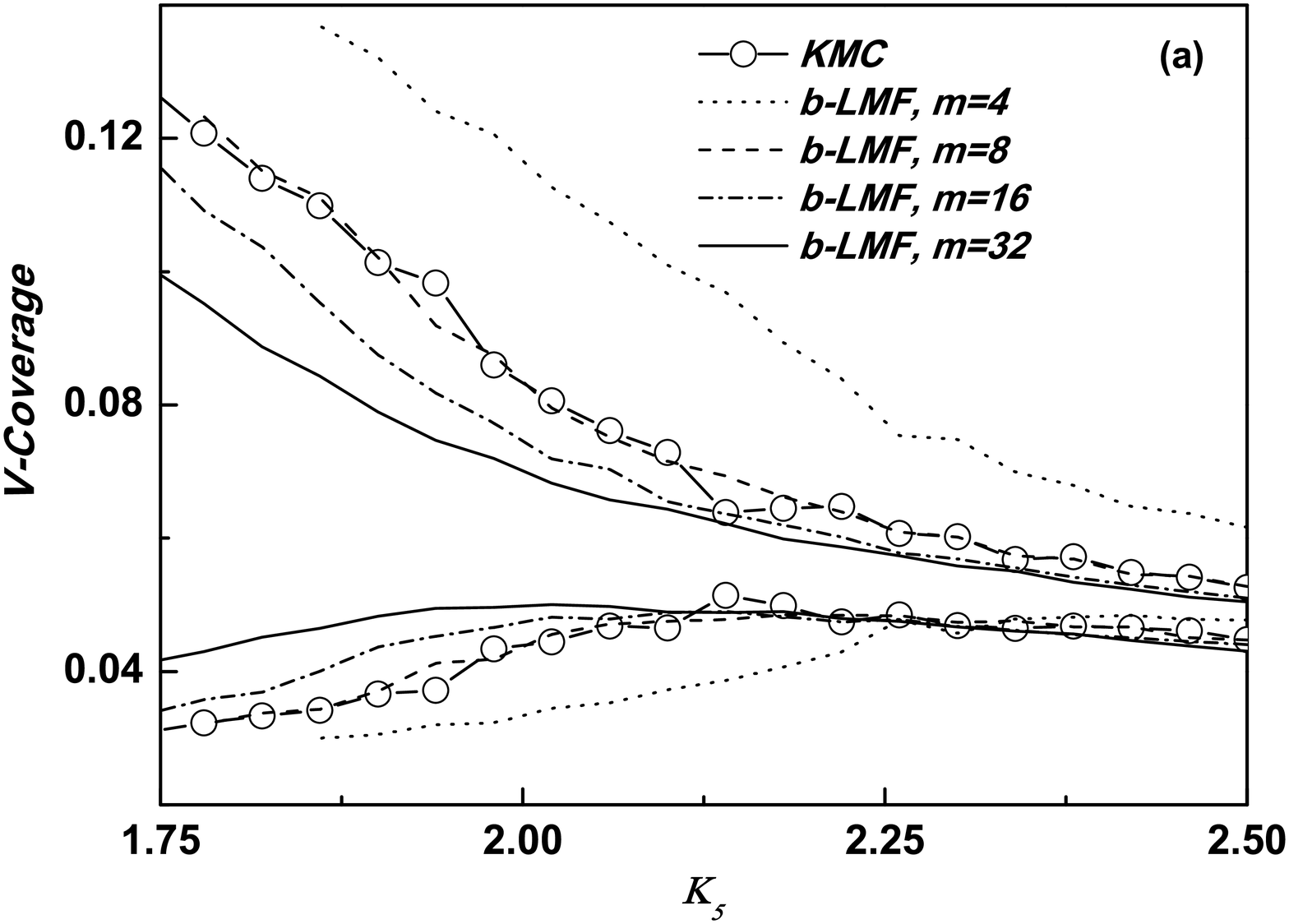}
\includegraphics [width=0.45\columnwidth,height=5cm]{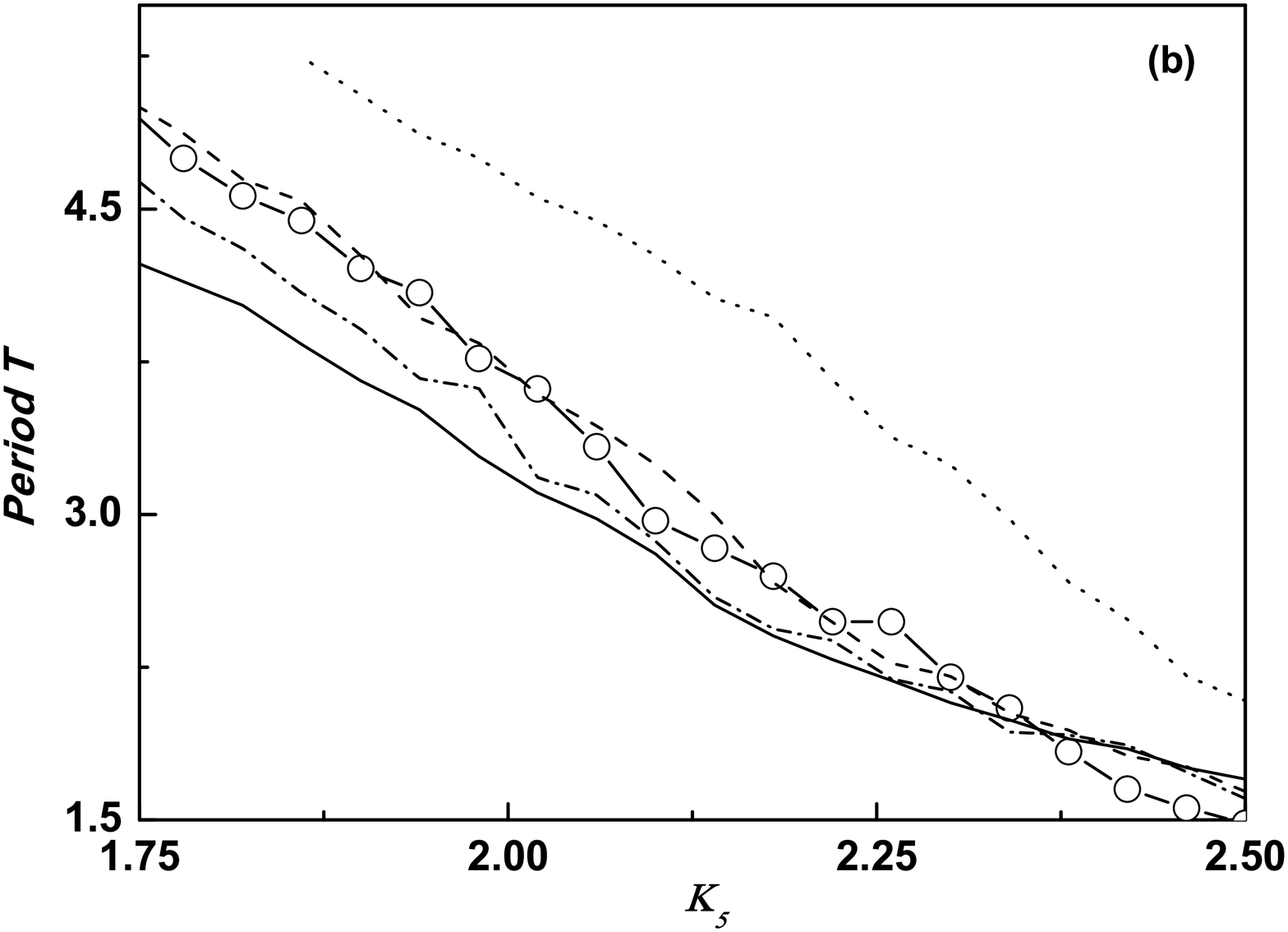}
\caption{ The oscillation range of (a) $V$-coverage and (b) period
obtained from the $b$-LMF CG approach with different coarse size
$m$. The diffusion constant is $D=10$ and the other parameters are
the same as in Fig. 2. \label{fig4}}
\end{center}
\end{figure}

In Fig.5, we have plotted the dependence of the turnover frequency
(TOF) as a function of time for different coarse size $m$ obtained
from the KMC, b-LMF and s-LMF. The TOF is defined as the occurrence
of the trimolecular reaction (5) per surface site per unit time.
Clearly, the b-LMF with $m=8$ matches the KMC quite well, while that
with smaller or larger $m$ may capture some qualitative features of
the TOF but with apparent quantitative differences. The s-LMF,
however, almost loses the temporal information associated with the
TOF.

\begin{figure}
\begin{center}
\includegraphics [width=0.9\columnwidth]{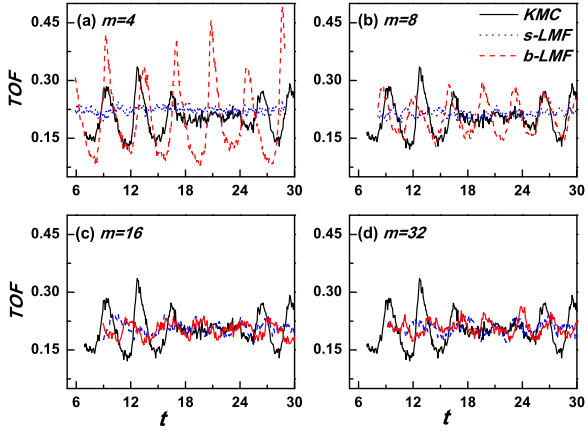}
\caption{ The dependence of TOF as a function of time on the coarse
cell size $m$ obtained from the KMC(solid-line), $b$-LMF(dash-line)
and $s$-LMF(dot-line), respectively. \label{fig6}}
\end{center}
\end{figure}

To further demonstrate this quantitatively, we introduce a deviation
coefficient for the oscillation range as follows. According to Fig.3
and Fig.4, each bifurcation diagram contains two branches, the upper
branch and the lower one corresponding to the averaged maximum and
minimum values of $v(t)$, respectively. As can be seen from the
figures, both branches obtained from the CG-KMC methods show
discrepancies with the KMC values. Denote the upper branch value of
$v$, obtained by CG-KMC, at a certain control parameter $K_5$ by
$v^u_k$, and that obtained by KMC by $v^u_{k0}$, then
$$\gamma^u_A=\frac{1}{2N_k}\sum_k \frac{\vert v^u_k-v^u_{k0} \vert}{v^u_{k0}} $$
measures the relative discrepancy of the upper branch, where $N_k$
is the number of control parameters used in the calculation.
Similarly, we can calculate the discrepancy of the lower branch
$\gamma^l_A$. In the present work, we have used $N_k=20$ points
inside the range $K_5 \in (1.75, 2.5)$ to obtain $\gamma^u_A$ and
$\gamma^l_A$. In Fig.(6a), the dependence of
$\gamma_A=(\gamma^u_A+\gamma^l_A)/2$ on the size $m$ of CG-cell is
shown, for the b-LMF with different diffusion constant $D$ and the
s-LMF with $D=10$. In Fig.(6b), the curves for $\gamma_T$, the
relative discrepancy in the period, are shown. Clearly, the s-LMF
method shows relatively large discrepancies, while the b-LMF works
much better. The s-LMF is even worse than the MFDE, shown by the
dash lines in Fig.6 for $D=10$. One notes that both $\gamma_A$ and
$\gamma_T$ exhibit a clear-cut minimum of about zero at $m=8$ for
large $D$ when b-LMF is used. We also note that if $D$ is small, say
$D=1$ here, the b-LMF also fails. This is not surprising because CG
method which assumes well-mixing in a CG-cell should not work if
diffusion is too slow.

In the above results, we see that the CG results for small $m$ do
not match the results of KMC, even for the b-LMF. This is in
contrast to the CG-KMC methods used by Vlachos et.al to account for
the dynamics of 2D lattice gas Ising model. Note that for the Ising
model, one mainly considered the equilibrium states. For the
Brusselator model considered here, however, we want to reproduce the
temporal oscillation behavior. To reproduce the oscillation features
on the whole surface, the temporal correlation of the time series
must be properly maintained during the CG procedure. When we perform
the CG procedure by dividing the lattice into CG cells, we are
dealing with $N^c\times N^c$ coupled CG oscillators, where $N^c=N/m$
is the size of the CG lattice. The coupling between these CG
oscillators are realized by the diffusion and the boundary
correlation considered in the b-LMF. If the CG cell is too small,
however, the time-correlation inside each cell will be lost due to
strong fluctuations and the time evolution of $\eta^\phi(t)$ cannot
be viewed as an oscillation. As discussed by P.Gaspard, a minimum
number of well-mixed molecules is required to produce correlated
oscillations, such that the auto-correlation time of the time series
is not smaller than $T/2\pi$\cite{Gaspard,Zhadnov2001}. Therefore, it seems that
a seamless CG approach to reproduce temporal dissipative structures
like chemical oscillation is a large challenge. On the other hand,
for any CG method within LMF scheme to work well, the scale of a CG
cell should not be larger than the diffusion length, as emphasized
by Mikhailov and others\cite{Mik0,Vlachos2}. The compromise between
these two factors, i.e., to keep time autocorrelation and to be
smaller than the diffusion length, may be the reason of appearance
of an optimal $m$ for the b-LMF approach. We note that this
reasoning is not applicable to the s-LMF, for which the
discrepancies monotonically decrease with increasing $m$, since the
s-LMF does not work for the present system.

\begin{figure}
\begin{center}
\includegraphics [width=0.45\columnwidth,height=5cm]{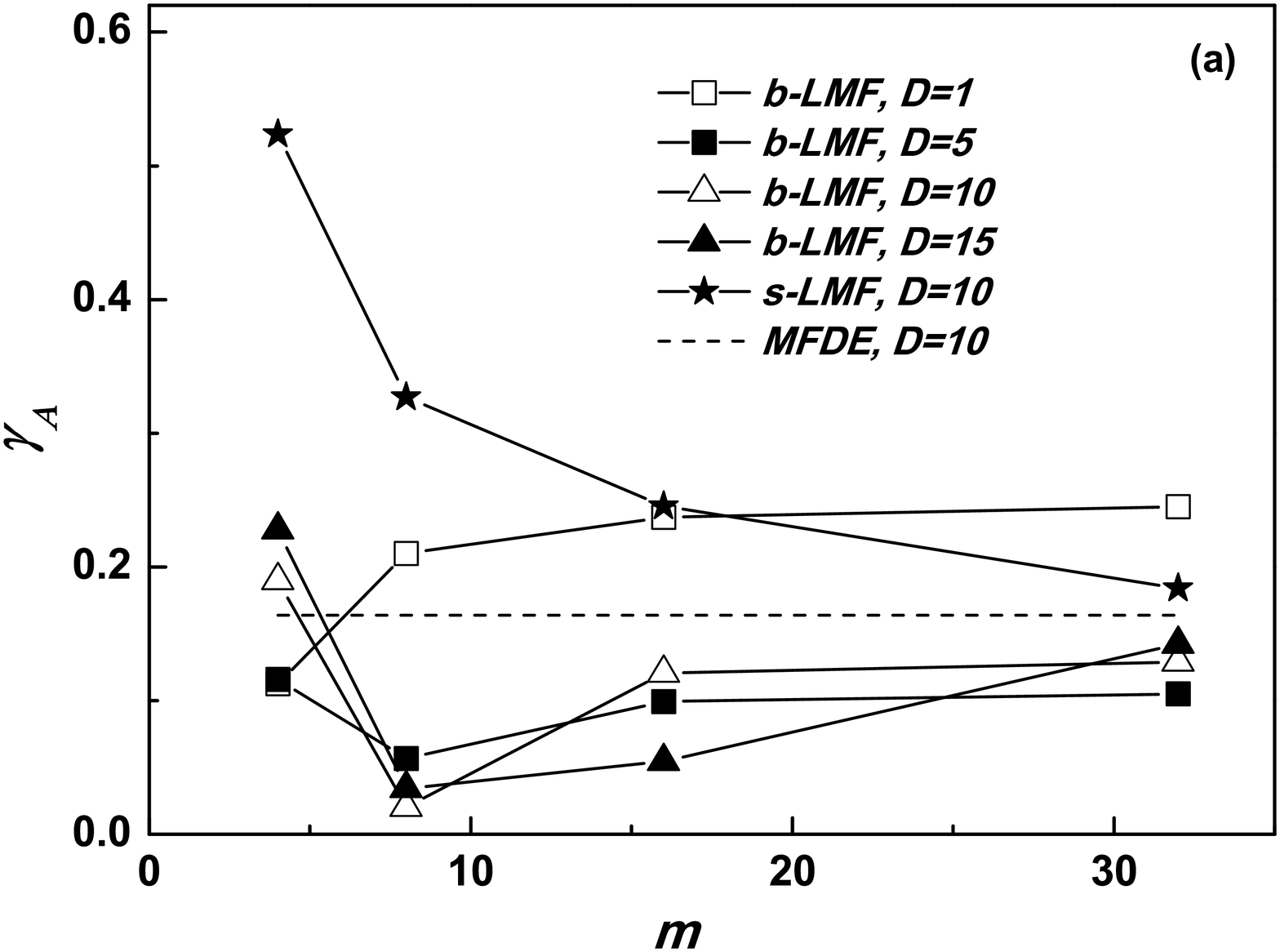}
\includegraphics [width=0.45\columnwidth,height=5cm]{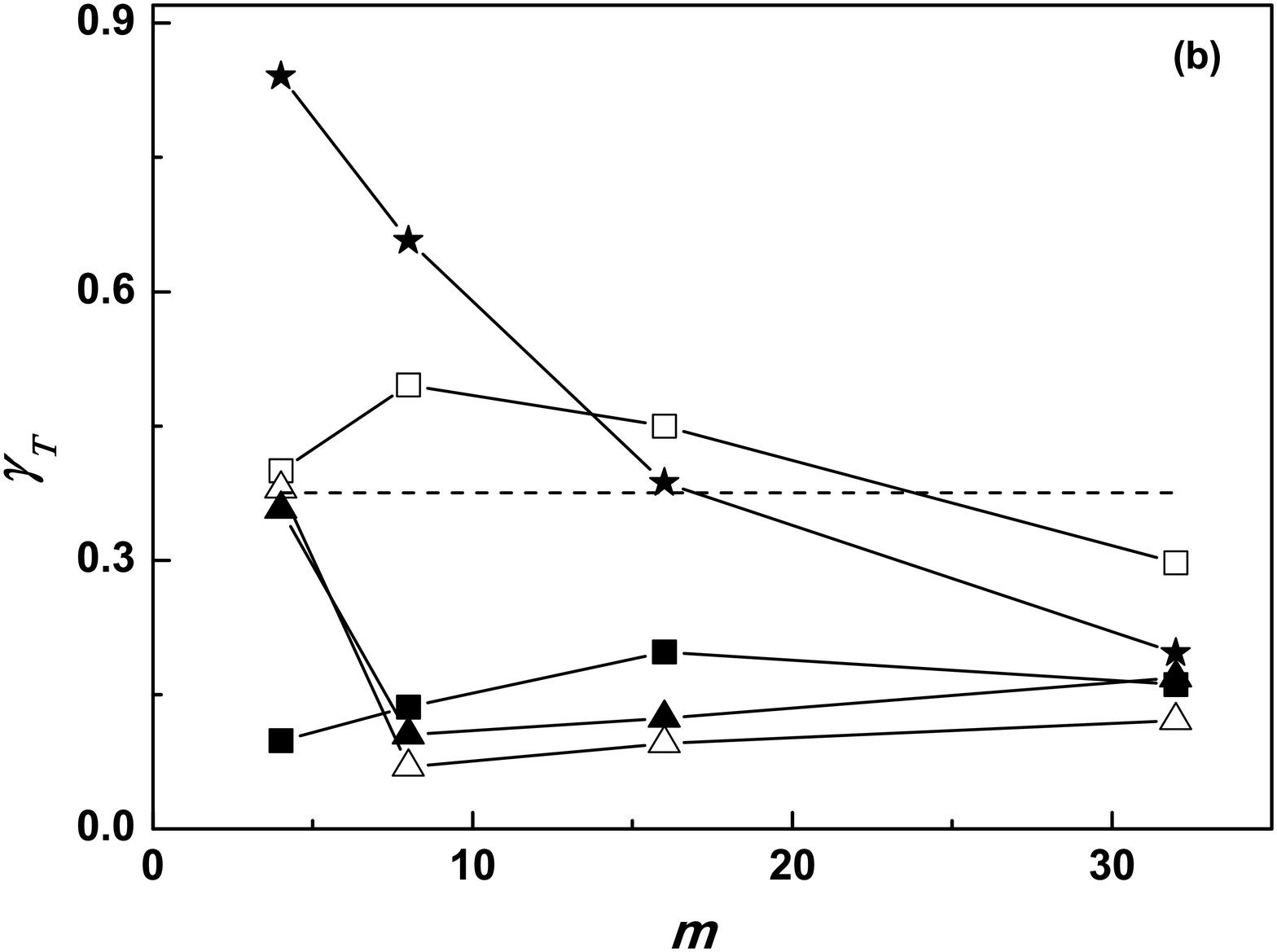}
\caption{ The dependence of the deviation coefficient (a) for the
oscillation amplitude $\gamma_A$ and (b) oscillation period
$\gamma_T$ on the coarse cell size $m$ obtained from the $s$-LMF and
$b$-LMF methods. Parameter are the same as in Fig. 2 and $N = 256$.
\label{fig5}}
\end{center}
\end{figure}

\section{Conclusions} \label{sec4}
In summary, we have tried to apply a CG-KMC approach to simulate the
oscillation behavior on the surface lattice-gas Brusselator model.
Owing to the correlations between adjacent cells resulted from the
nonlinear trimolecular reaction, the CG approach based on simple LMF
approximation almost fails. By properly taking into account the
boundary corrections, we have introduced a so-called b-LMF CG-KMC
approach, which can reproduce the microscopic KMC results quite
well, given that the diffusion is not too slow and the CG cell size
is optimally chosen. Our work thus unravels the very role of
reaction correlations which should be carefully considered in any CG
approach and mesoscopic modeling for nonequilibrium spatiotemporal
dynamics at nanoscales.

\begin{acknowledgments}
This work was supported by the National Natural Science Foundation
of China under Grant No.20933006 and No.20873130.
\end{acknowledgments}


\begin{thebibliography}{32}

\bibitem{A.S.M and G.E} {\ A. S. Mikhailov and G. Ertl,} { Chem. Phys. Chem.} \textbf{10}, 86 (2009).
\bibitem{Mik4}{\ M. Hildebrand, M. Ipsen, A. S. Mikhailov and G. Ertl,} {New J. Phys.} \textbf{54}, 61(2003).
\bibitem{Mik5}{\ Y. De Decker and A. S. Mikhailov,} {J. Phys. Chem. B} \textbf{108}, 14759(2004).
\bibitem{Progress-165} {\ Y. De Decker and A. S. Mikhailov,} { Prog. Theo. Phys. Supp.} \textbf{165}, 119 (2006).
\bibitem{Science-293} {\ C. Sachs, M. Hildebrand, S. Volkening, J. Wintterlin and G. Ertl,} { Science} \textbf{293}, 1635
(2001).
\bibitem{Nature-V390} {\ T. Zambelli, J. V. Barth, J. Wintterlin and G. Ertl,} { Nature} \textbf{390}, 495 (1997).
\bibitem{PRL-V65} {\ S. Jakubith, H. H. Rotermund, W. Engel, A. von Oertzen and G. Ertl,} { Phys. Rev. Lett.} \textbf{65},
3013
(1990).
\bibitem{Science-278} {\ J. Wintterlin, S. V\"{o}lkening, T. V. W. Janssens, T. Zambelli and G. Ertl,} { Science}
\textbf{278},
1931(1997).
\bibitem{JCS} {\ M. Gruyters, D. A.King,} { J. Chem. Soc.} \textbf{93}, 2947(1997).
\bibitem{Book} {\ M. M. Slinko, N. I. Jaeger,} {\em Oscillatory Heterogeneous Catalytic Systems}, Elsevier, Amsterdam, 1994.

\bibitem{MF1} {\ E. E. Mola, I. M. Irurzun, J. L. Vicente and D. A. King,} { Surf. Rev. Lett.} \textbf{10}, 23(2003).
\bibitem{MF2} {\ F. Sch\"{u}th, B. E. Henry and L. D. Schmidt,} { Adv. Catal.} \textbf{39}, 51 (1993).

\bibitem{KMC1} {\ A. Provata and V. K. Noussiou,} { Phys. Rev. E} \textbf{72}, 066108 (2005).
\bibitem{KMC2} {\ V. P. Zhdanov and T. Matsushima,} { Surf. Sci.} \textbf{583} 253(2005).
\bibitem{KMC3} {\ V. P. Zhdanov,} { Surf. Sci. Rep.} \textbf{45}, 231 (2002).
\bibitem{KMC4} {\ V. P. Zhdanov and B. Kasemo,} { Surf. Sci. Rep.} \textbf{39}, 25(2000).
\bibitem{KMC5} {\ V. P. Zhdanov and T. Matsushima,} {Phys. Rev. Lett} \textbf{98}, 036101(2007).
\bibitem{KMC6} {\ V. P. Zhdanov,} {J. Chem. Phys.} \textbf{126}, 074706(2007).

\bibitem{Mik0} {M. Hildebrand and A. S. Mikhailov,} {J. Phys. Chem.} \textbf{100}, 19089(1996).
\bibitem{Mik1} {M. Hildebrand, A. S. Mikhailov and G. Ertl,} {Phys. Rev. E} \textbf{58}, 5483(1998).
\bibitem{Mik2} {M. Hildebrand, A. S. Mikhailov and G. Ertl,} {Phys. Rev. Lett.} \textbf{81}, 2602(1998).
\bibitem{Mik3} {M. Hildebrand,} {Chaos} \textbf{12}, 144(2002).
\bibitem{Mik6} {M. Hildebran and A. S. Mikhailov,} {J. Stat. Phys.} \textbf{101}, 599(2000).
\bibitem{Mik7} {A. S. Mikhailov, M. Hildebrand and G. Ertl,} {in \em Coherent Structures in Complex Systems}, {Springer, New York, 252(2001)}

\bibitem{Vlachos0} {\ A. Chatterjee and D. G. Valchos,} { J. Chem. Phys.} \textbf{124}, 064110 (2006).
\bibitem{Vlachos1} {\ M. A. Katsoulakis and D. G. Vlachos,} { J. Chem. Phys.} \textbf{119}, 9412 (2003).
\bibitem{Vlachos2} {\ A. Chatterjee and D. G. Vlachos,} { J. Chem. Phys.} \textbf{121}, 11420 (2004).
\bibitem{Vlachos3} {\ M. A. Katsoulakis, A. J. Majda and D. G. Vlachos,} { J. Comput. Phys.} \textbf{186}, 250 (2003).
\bibitem{Vlachos4} {\ M. A. Katsoulakis, A. J. Majda and D. G. Vlachos,} { Proc. Natl. Acad. Sci.} \textbf{100}, 782 (2003).
\bibitem{Vlachos6} {\ A. Chatterjee and D. G. Vlachos,} { Chem. Eng. Sci.} \textbf{62}, 4852(2007).
\bibitem{Vlachos7} {\ S. D. Collins, A. Chatterjee and D. G. Vlachos,} { J. Chem. Phys.} \textbf{129}, 184101(2008).
\bibitem{Vlachos8} {\ S. D. Collins, M. Stamatakis and D. G. Vlachos,} { BMC Bioinformatics} \textbf{11}, 218(2010).

\bibitem{G Ertl} {\ R. Imbihl and G. Ertl,} { Chem. Rev.} \textbf{95}, 697 (1995).
\bibitem{Review of KMC} {A. Chatterjee and D. G. Vlachos,} { J. Comput.-Aided Mater Des} \textbf{14}, 253 (2007).
\bibitem{Zhadnov2001} {\ V.P. Zhdanov,} { Phys. Chem. Chem. Phys.} \textbf{3}, 1432(2001).

\bibitem{PNAS} {D. Fange, O. G. Berg, P. Sj\"{o}berg and J. Elf,} {Proc. Natl. Acad. Sci.}, \textbf{107}, 19820(2010).

\bibitem{Gaspard} {\ P. Gaspard,} { J. Chem. Phys.} \textbf{117}, 8905 (2002).

\end{thebibliography}

\end{document}